%% file: document.tex
\documentclass[12pt]{article}
\usepackage{graphicx}
\usepackage[dvips]{hyperref}
\newcommand\pubnumber{XXXX-XX-XX}
\newcommand\pubdate{September 21, 2011}%\today}

\def\muenster{European Organization for Nuclear Research (CERN), Geneva, Switzerland and \newline 
Institut f\"ur Kernphysik, Westf\"alische Wilhelms-Universit\"at M\"unster, Germany \newline
eva.sicking@cern.ch}

\def\Title#1{\begin{center} {\Large #1 } \end{center}}
\def\Author#1{\begin{center}{ \sc #1} \end{center}}
\def\Address#1{\begin{center}{ \it #1} \end{center}}

\newcommand\pubblock{\rightline{\begin{tabular}{l} \pubnumber\\
         \pubdate  \end{tabular}}}
\newenvironment{Abstract}{\begin{quotation}  }{\end{quotation}}
\newenvironment{Presented}{\begin{quotation} \begin{center} 
             PRESENTED AT\end{center}\bigskip 
      \begin{center}\begin{large}}{\end{large}\end{center} \end{quotation}}

\input econfmacros.tex
\begin{document}
\begin{titlepage}
\pubblock

\vfill
\Title{Minimum Bias Measurements with ALICE at the LHC}
\vfill
\Author{ Eva Sicking on behalf of the ALICE Collaboration}%\support}
\Address{\muenster}
\vfill
\begin{Abstract}
ALICE (A Large Ion Collider Experiment) is one of the seven experiments at the the Large Hadron Collider (LHC) at CERN, Geneva, Switzerland. ALICE is especially designed for heavy-ion collisions but it also operates a rich proton-proton (pp) program. ALICE has collected pp collision data at $\sqrt{s}=$ 0.9, 2.36, 2.76, and 7 TeV and lead-lead (Pb--Pb) collision data at $\sqrt{s_{\mathrm{NN}}}=$2.76 TeV.\\
Here, we report minimum bias measurements obtained until the end of 2010: the results include measurements of charged-particle pseudorapidity, multiplicity and transverse momentum distributions. Also, the two-pion Bose-Einstein correlation and the measurement of antiproton-to-proton ratio will be discussed. Furthermore, results on the production of identified particles including strange particles will be shown as well as first results from the first Pb--Pb run at the LHC.
\end{Abstract}
\vfill
\begin{Presented}
MPI@LHC 2010: $2^{\mathrm{nd}}$ International Workshop on Multiple Partonic Interactions at the LHC\\
Glasgow, $29^{\mathrm{th}}$ of November to the $3^{\mathrm{rd}}$ of December 2010.
\end{Presented}
\vfill
\end{titlepage}
\def\thefootnote{\fnsymbol{footnote}}
\setcounter{footnote}{0}

\section{Introduction}
ALICE \cite{Aamodt:2008zz} is the experiment at the LHC that was especially designed for heavy-ion collisions. It also has a rich pp program that benefits from ALICE's low momentum sensitivity due to its small material budget and its comparatively low magnetic field. 
ALICE has a good primary vertex resolution in $x,y$ direction of $100\, \mu$m in pp and $10\,\mu$m in Pb--Pb collision events as well as a good secondary vertex resolution. Furthermore, ALICE has excellent particle identification capabilities using various methods, e.g.\ specific energy loss, transition radiation, time of flight, and Cherenkov radiation. \\
In the presented analyses, the trigger and reconstruction efficiencies and the remaining contamination from secondary particles were estimated via simulations within the AliRoot simulation and analysis framework \cite{:2005ji}: PHOJET \cite{Engel:1995sb} and PYTHIA with different tunes \cite{Sjostrand:2006za, Skands:2010ak, Moraes:2009zz, Albrow:2006rt}  were used as event generators for pp collisions. In the Pb--Pb analyses, HIJING \cite{Wang:1991us}, DPMJET \cite{Roesler:2000he}, and THERMINATOR \cite{Kisiel:2005hn} were used. The detector response was modeled using GEANT3 \cite{:1994zzo}.

\section{Pseudorapidity density and multiplicity}
ALICE has measured the charged particle pseudorapidity density and multiplicity distribution in pp collisions for inelastic (INEL) collisions, non-single-diffractive (NSD) collisions, and for events with at least one track in $|\eta|<1$
\begin{figure}[h]
\begin{minipage}[b]{0.4\linewidth}
\centering
\includegraphics[width=\textwidth]{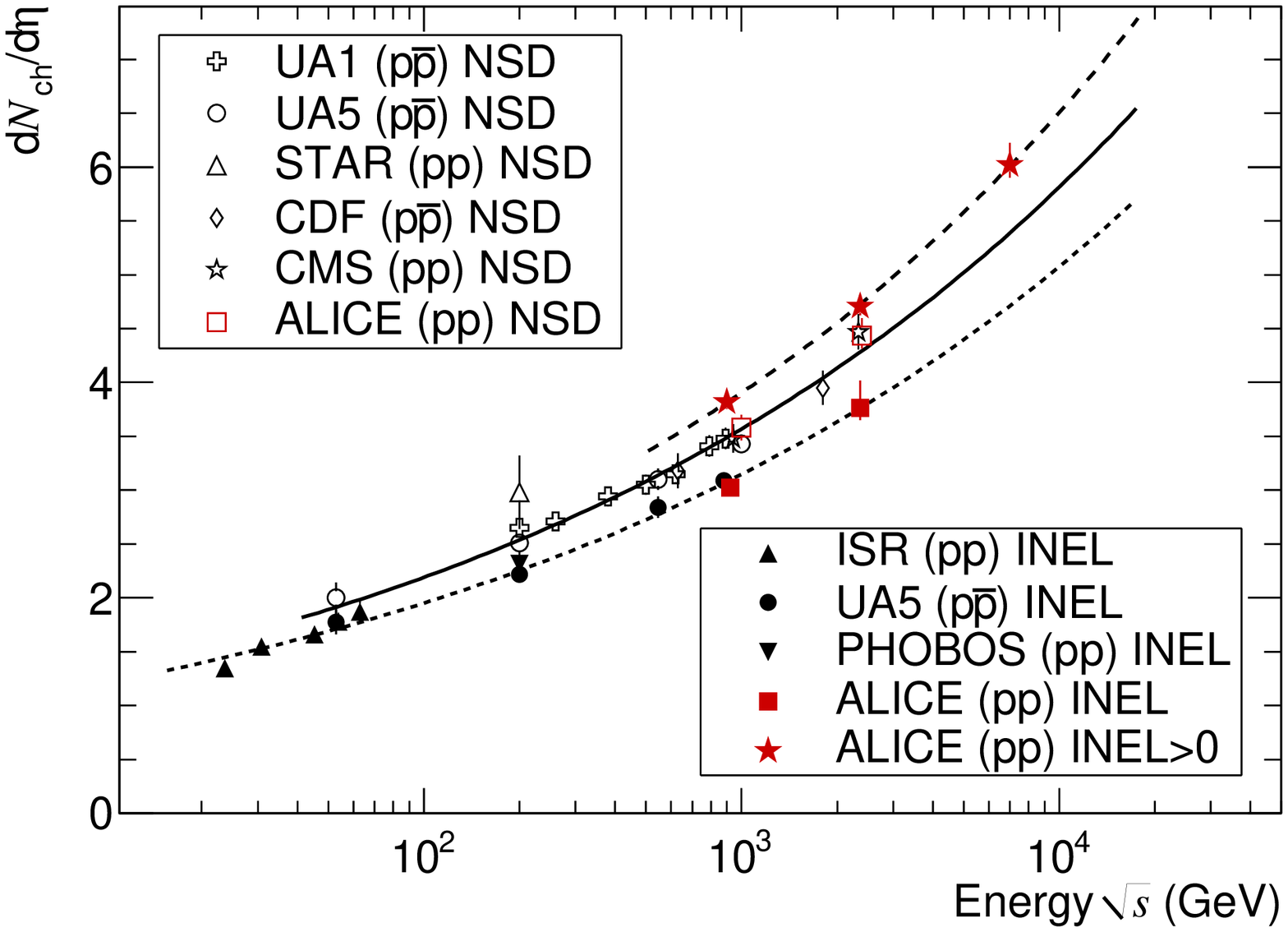}
\end{minipage}
\vspace{-0.5cm}
\begin{minipage}[b]{0.4\linewidth}
\flushright
\includegraphics[width=\textwidth]{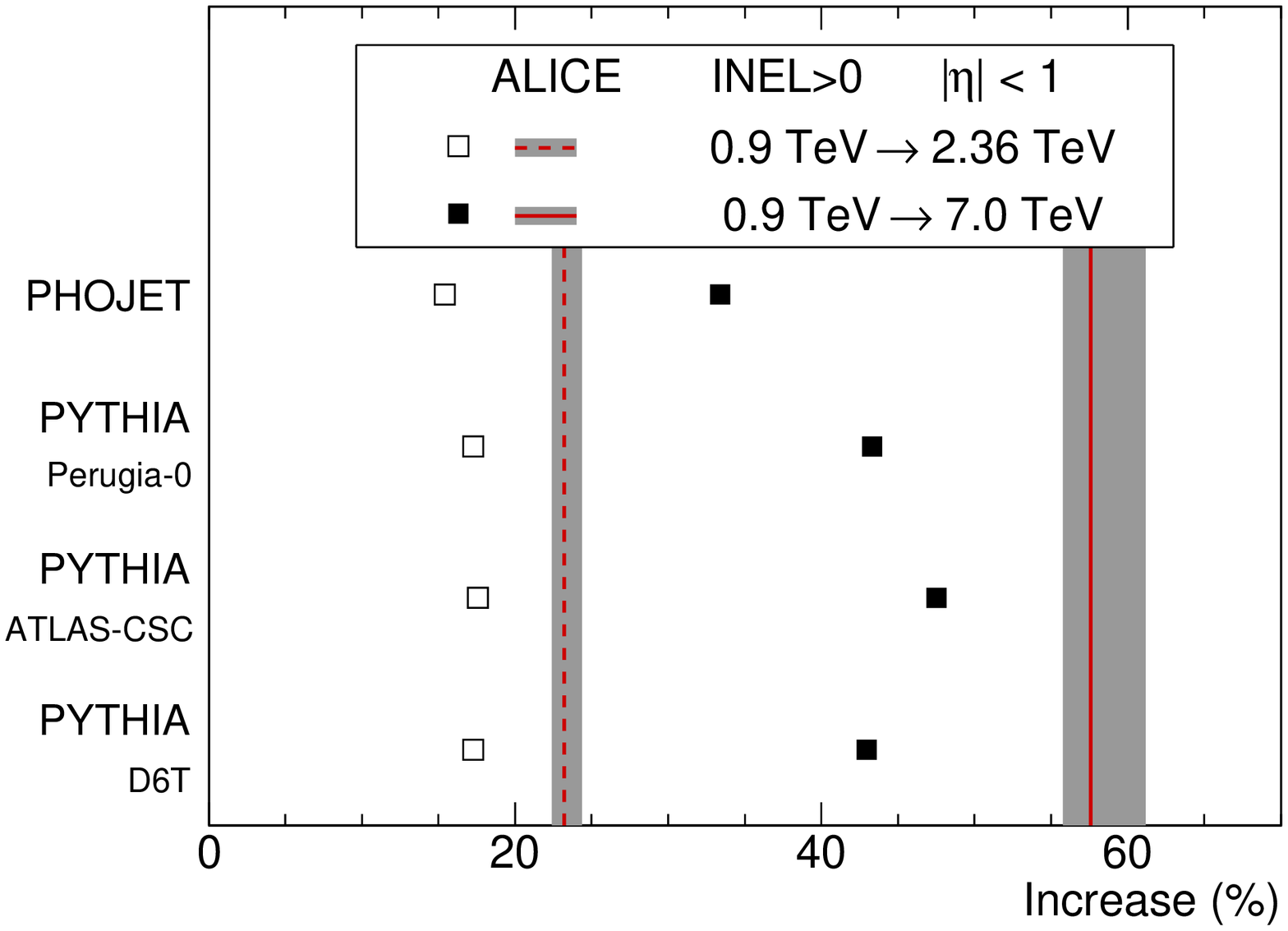}
\end{minipage}
\begin{minipage}[t]{0.48\linewidth}
\centering
\caption{Charge particle pseudorapidity density in the central pseudorapidity region as function of center-of-mass energy measured by different experiments. Presented are results for INEL and NSD and $\mathrm{INEL}>0_{|\eta|<1}$ event classes \cite{Aamodt:2010pp}.} 
\label{fig:eta1}
\end{minipage}
\hspace{0.6cm}
\begin{minipage}[t]{0.48\linewidth}
\centering
\caption{Relative increase of the charged particle pseudorapidity density for the $\mathrm{INEL}>0_{|\eta|<1}$ event class between $\sqrt{s}=0.9$ TeV and $\sqrt{s}=2.36$ TeV and between $\sqrt{s}=0.9$ TeV and $\sqrt{s}=7.0$~TeV, respectively \cite{Aamodt:2010pp}. Open and solid symbols refer to expectations from MC generators using different tunes.}
\label{fig:eta2}
\end{minipage}
\end{figure}
 ($\mathrm{INEL}~>~0_{|\eta|<1}$) \cite{:2009dt, Aamodt:2010ft,Aamodt:2010pp}. 
The measurements were performed at $\sqrt{s}=$ 900~GeV, 2.36~TeV and 7~TeV. 
The results for the central rapidity region measured by ALICE and other experiments are presented in Figure \ref{fig:eta1}.
The increase with the center-of-mass energy is well described by a power law.
The results of ALICE and CMS are in good agreement.\\
Figure \ref{fig:eta2} presents the relative increase of the charged particle pseudorapidity density for the $\mathrm{INEL}>0_{|\eta|<1}$ event class
between $\sqrt{s}=0.9$ TeV and $\sqrt{s}=2.36$ TeV and between $\sqrt{s}=0.9$ TeV and $\sqrt{s}=7.0$ TeV respectively for different MC models in comparison to ALICE measurements. The increase measured in ALICE is significantly larger than predicted by the event generators.\\
Figure \ref{fig:mult} shows the measured charged particle multiplicity distributions in $|\eta|<1$ for the $\mathrm{INEL}>0_{|\eta|<1}$ event class. Results for three different collision energies are presented. A negative binomial distribution describes the shape of the distributions well. The right hand side of Figure \ref{fig:mult} shows, exemplarily at $\sqrt{s} = 7$ TeV, that none of the presented event generators is able to reproduce the shape and the tail of the multiplicity distribution. Solely, the PYTHIA tune ATLAS-CSC \cite{Moraes:2009zz, Albrow:2006rt} reproduces the data. However, this model fails at low multiplicities.
\begin{figure}[ht]
\begin{minipage}[b]{0.38\linewidth}
\centering
\includegraphics[width=\textwidth]{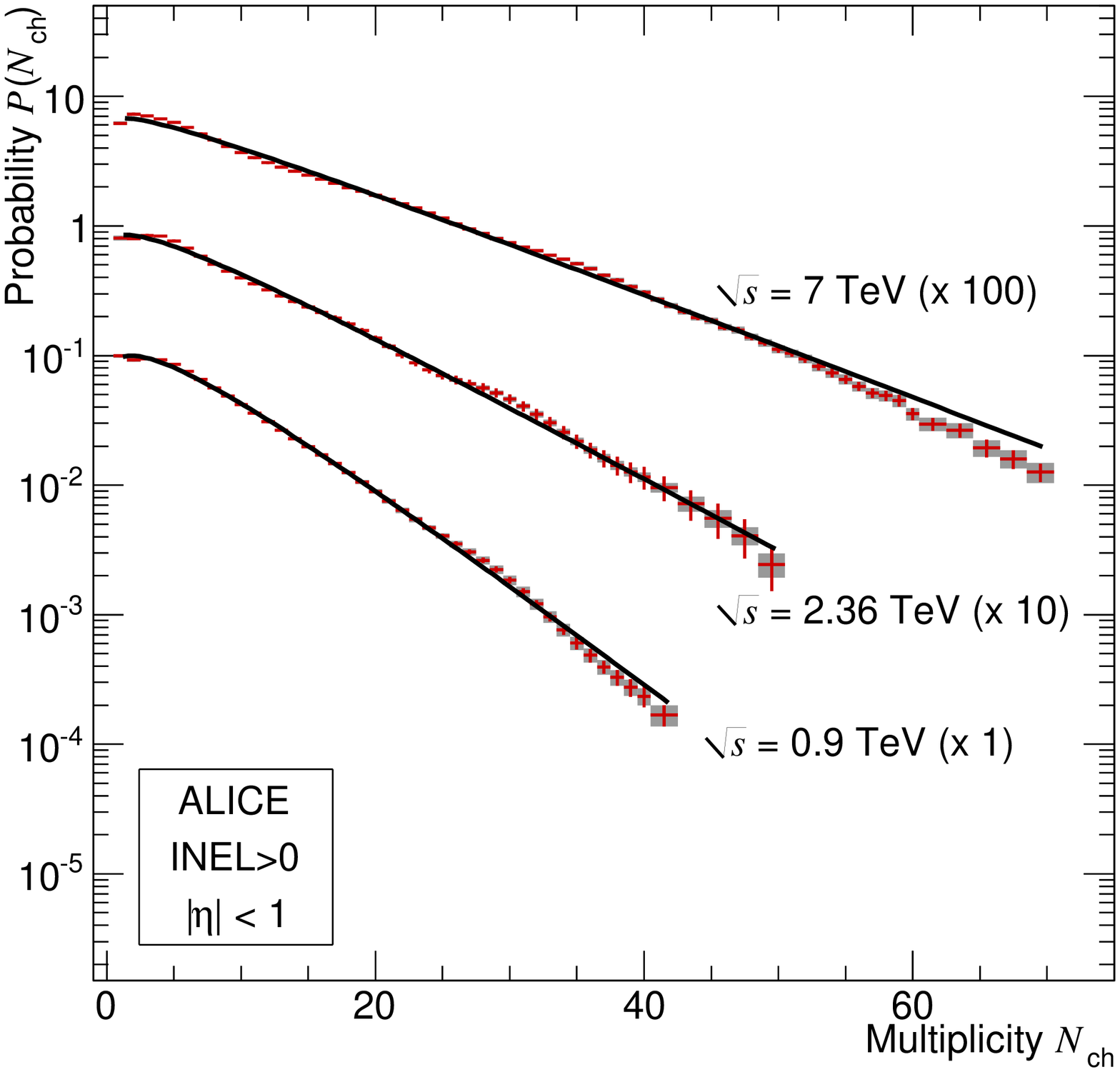}
\end{minipage}
\hspace{1.2cm}
\begin{minipage}[b]{0.38\linewidth}
\flushright
\includegraphics[width=\textwidth]{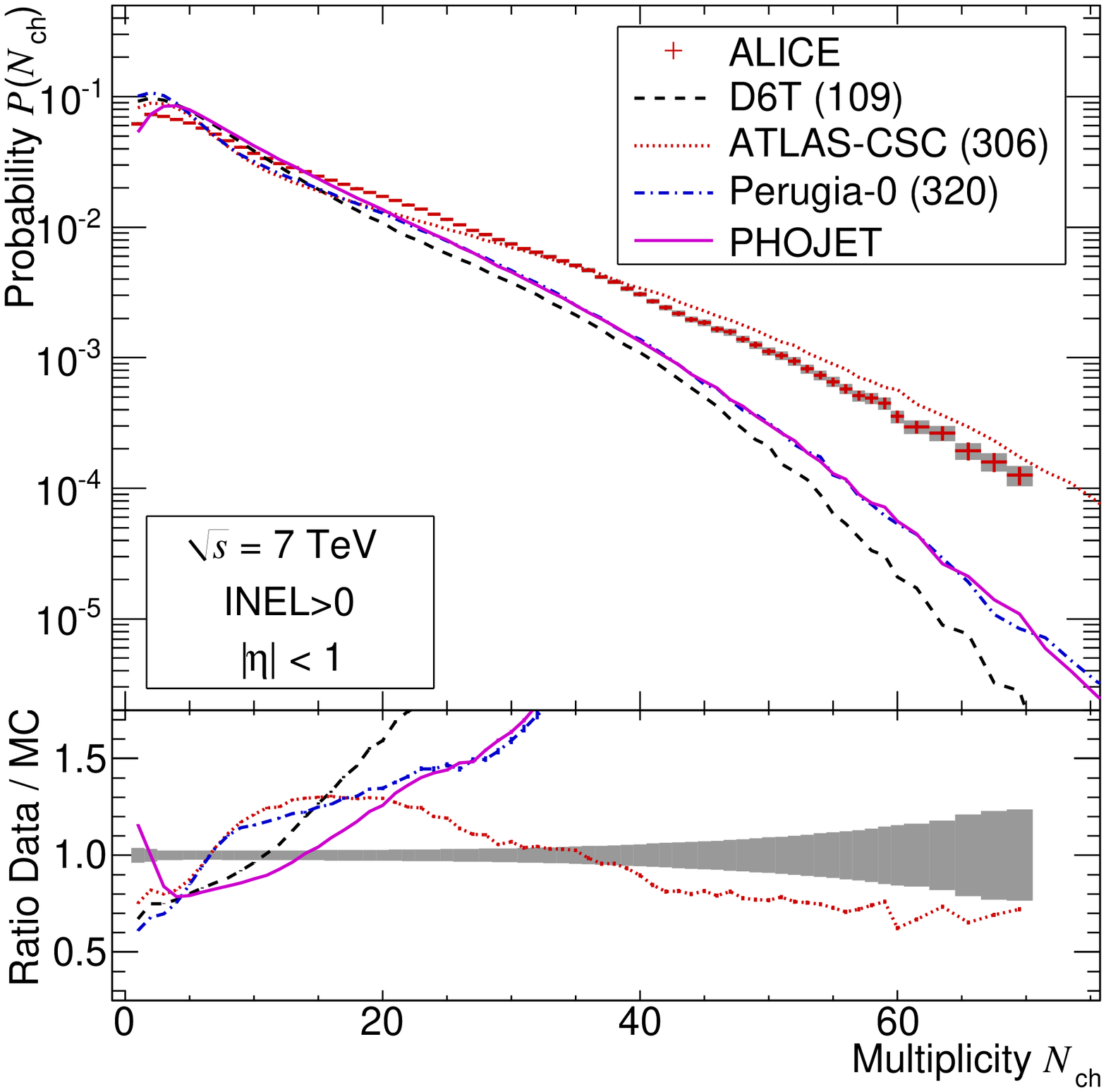}
\end{minipage}
\centering
\caption{Measured charged particle multiplicity distributions in $|\eta|<1$ for the $\mathrm{INEL}>0_{|\eta|<1}$ event class \cite{Aamodt:2010pp}. Left: Measured charged particle multiplicity distributions for three different energies are shown with non-binomial distribution fits. The results are scaled for better visibility. Right: Data at $\sqrt{s}=7$ TeV compared to MC models. In the lower part, the ratios between MC and data are shown.}
\label{fig:mult}
\end{figure}

\section{Transverse momentum distribution}
\begin{figure}[ht]
\begin{minipage}[b]{0.35\linewidth}
\centering
\includegraphics[width=\textwidth]{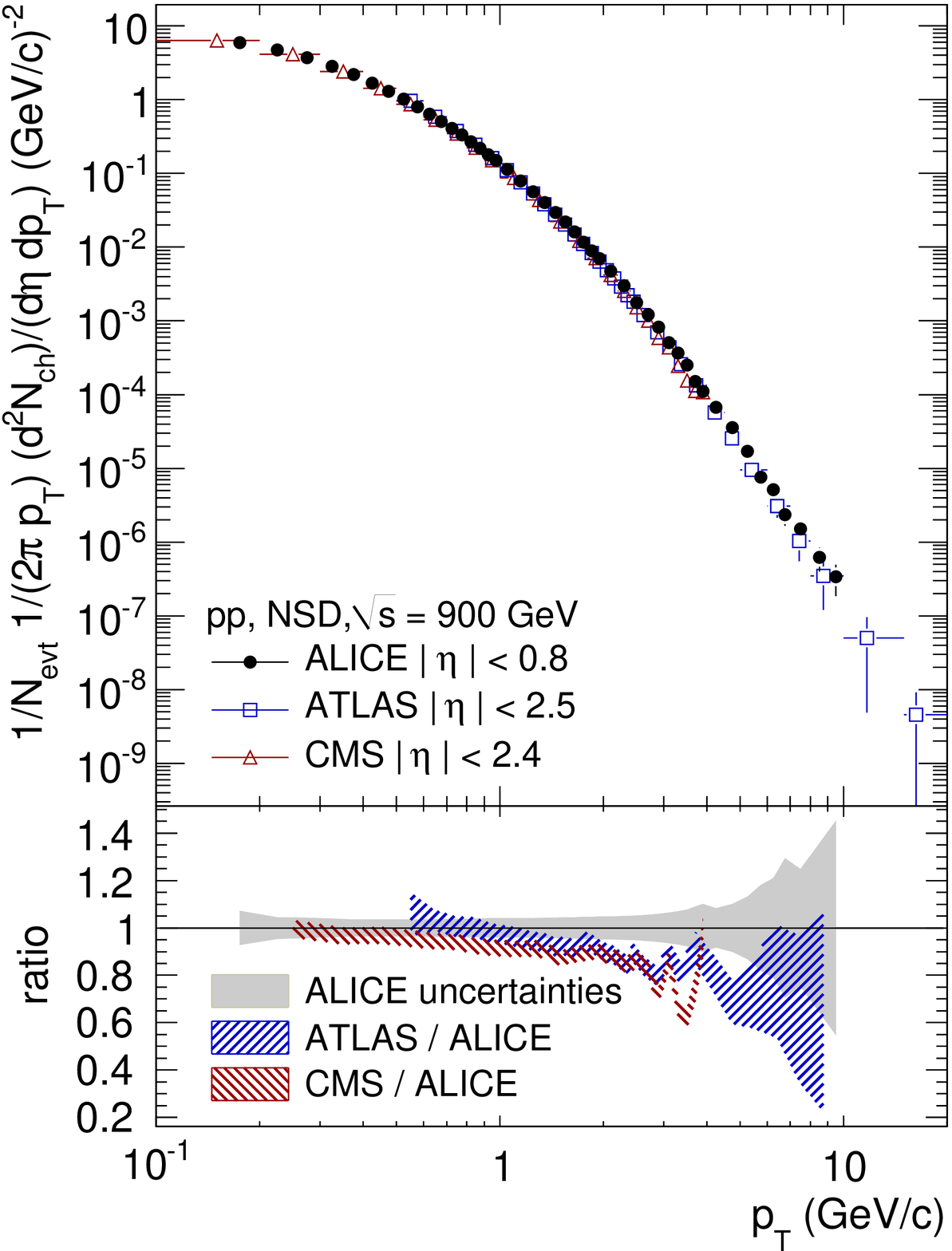}
\end{minipage}
%\hspace{1.2cm}
\vspace{-0.5cm}
\begin{minipage}[b]{0.35\linewidth}
%\flushright
\includegraphics[width=\textwidth]{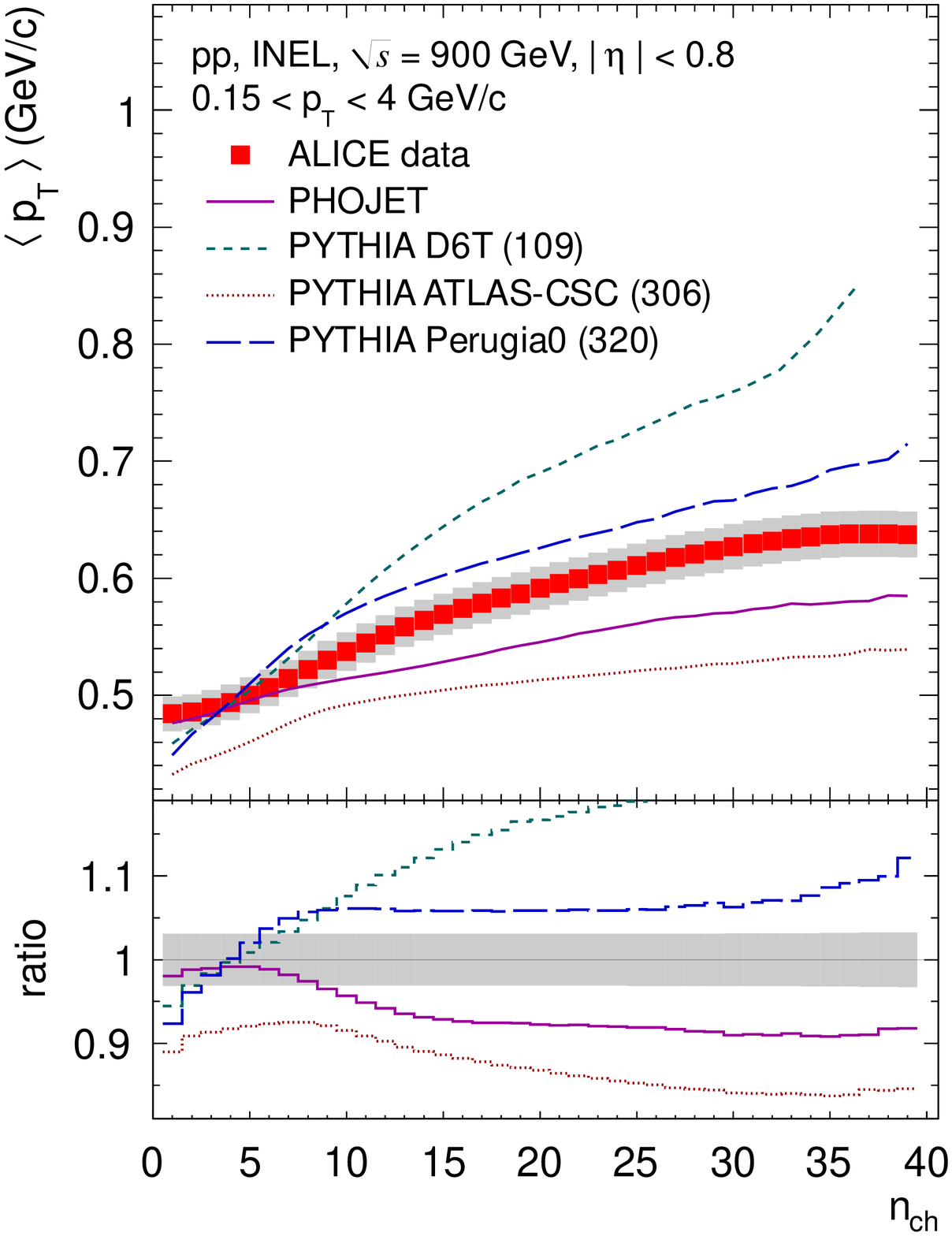}
\end{minipage}
\begin{minipage}[t]{0.45\linewidth}
\centering
\caption{Normalized differential primary particle yield in NSD collision averaged in $|\eta|<0.9$. ALICE data are compared to results of ATLAS and CMS (note the difference in the pseudorapidity coverage)~\cite{Aamodt:2010my}.}
\label{fig:pt1}
\end{minipage}
\vspace{-0.5cm}
\hspace{1.2cm}
\begin{minipage}[t]{0.45\linewidth}
\centering
\caption{The average transverse momentum of charged particles for $0.15 < p_{\mathrm{T}} < 4$ GeV/$c$ in inelastic pp events at 0.9 TeV as function of charged multiplicity in comparison to models \cite{Aamodt:2010my}.}
\label{fig:pt2}
\end{minipage}
\end{figure}

ALICE has measured the transverse momentum distribution $\mathrm{d}N_{\mathrm{ch}}/\mathrm{d}p_{\mathrm{T}}$ at $\sqrt{s}=0.9$~TeV for INEL and NSD events. In addition, the correlation between the average transverse momentum 
%in a collision event 
and charged multiplicity was studied  \cite{Aamodt:2010my}. \\
Figure \ref{fig:pt1} shows the normalized differential primary particle yield in NSD collisions averaged in $|\eta|<0.9$ compared to results from ATLAS and CMS. The higher yield at large $p_{\mathrm{T}}$ measured in ALICE is related to the difference in the acceptance for the different experiments: The spectrum gets harder with smaller rapidity windows around mid-rapidity.\\
Figure \ref{fig:pt2} presents the average transverse momentum of charged particles for $0.15 < p_{\mathrm{T}} < 4$ GeV/$c$ in inelastic pp events at 0.9 TeV as function of charged multiplicity in comparison to models. PYTHIA Perugia-0 and PHOJET are closest to the data. However, none of the models provide an exact description. For another transverse momentum range starting at a higher value, $0.5 < p_{\mathrm{T}} < 4$ GeV/$c$, PYTHIA Perugia-0 reproduces the data with good agreement (plot not shown). 

\section{Two-pion Bose-Einstein correlations}
The spatial extent of an emitting source can be measured via its Bose-Einstein enhanced correlations of identical pion pairs at low momentum differences. 
The two-particle correlation function is defined as $C({\bf q})= A({\bf q})/B({\bf q})$, where $\bf{q}=\bf{p_2}-\bf{p_1}$ is the four momentum difference of the two tracks of the pion pairs, $A(\bf{q})$ is the measured distribution of the pair momentum difference, and $B(\bf{q})$ is a similar distribution of pairs with pions from different events (mixed events) \cite{Aamodt:2010jj}.\\ 
ALICE has studied the two-particle correlation function $C({\bf q})$ in pp collisions at 0.9~TeV \cite{Aamodt:2010jj}. $C(\bf{q})$ was measured as function of the charged particle multiplicity and as function of the pair momentum $k_{\mathrm{T}} = |{\bf p_{\mathrm{T},1}}+{\bf p_{\mathrm{T},2}}|/2$. By fitting the correlation function with a Gaussian, the source size of the emitting system $R_{\mathrm{inv}}$ can be extracted.
%\vspace{-0.2cm}
\begin{figure}[ht]
\begin{minipage}[b]{0.39\linewidth}
\centering
\includegraphics[width=\textwidth]{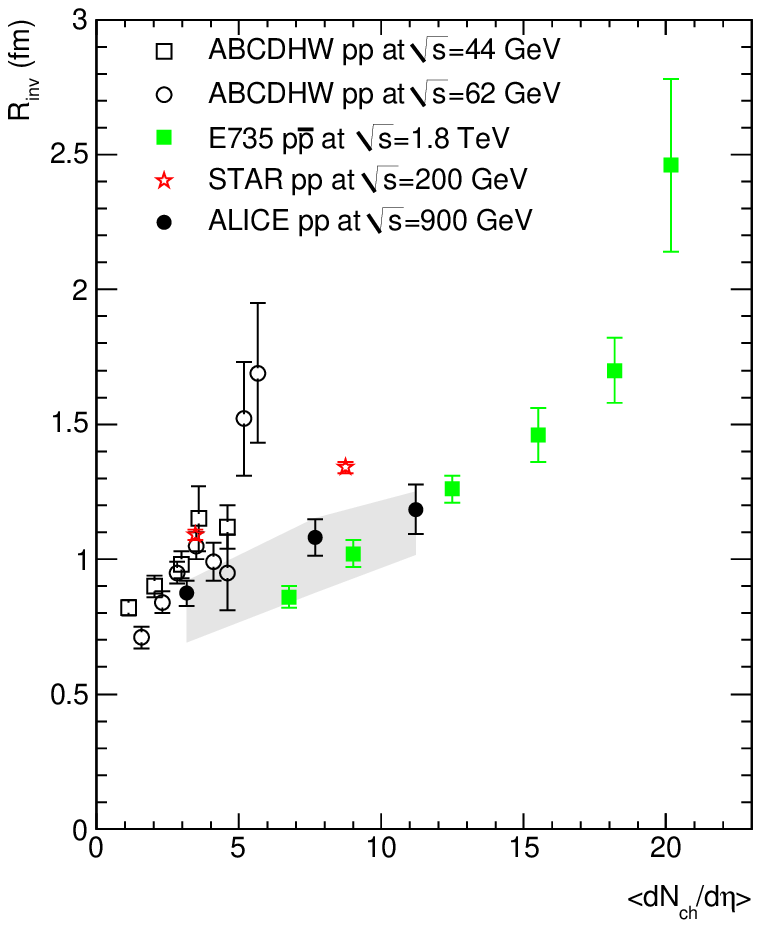}
\end{minipage}
\vspace{-0.3cm}
\begin{minipage}[b]{0.375\linewidth}
\includegraphics[width=\textwidth]{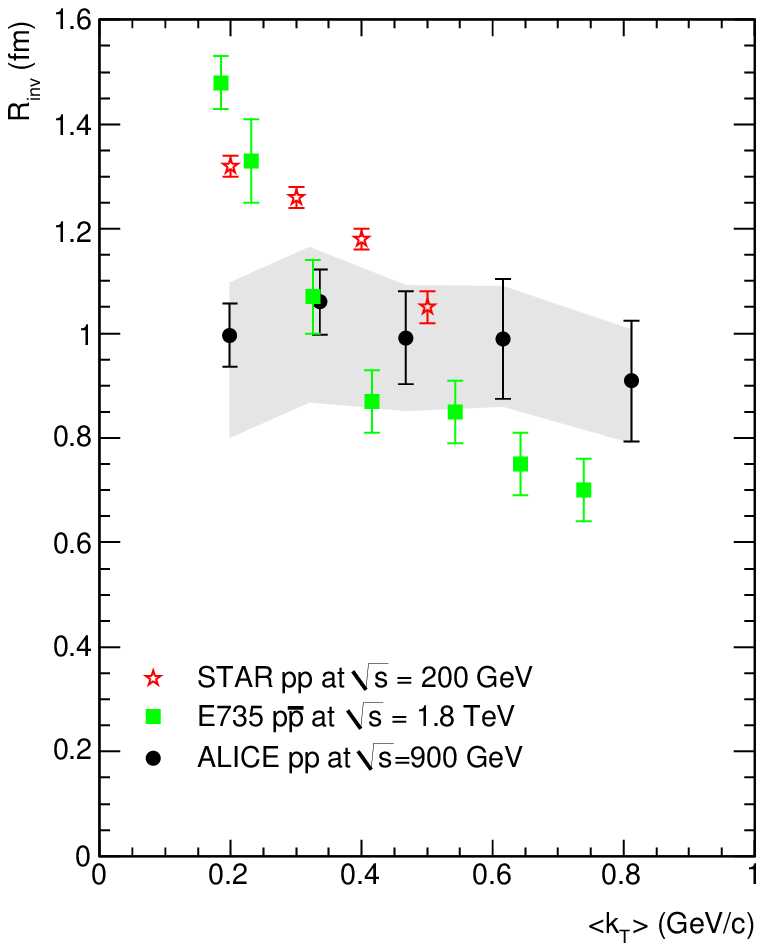}
\end{minipage}
\vspace{-0.3cm}
\begin{minipage}[t]{0.45\linewidth}
\centering
\caption{One dimensional Gaussian HBT radius $R_{\mathrm{inv}}$ in pp collisions at $\sqrt{s}=0.9$ TeV as function of charged particle multiplicity \cite{Aamodt:2010jj}.}
\label{fig:twopion}
\end{minipage}
\hspace{1.2cm}
\begin{minipage}[t]{0.45\linewidth}
\centering
\caption{One dimensional Gaussian HBT radius $R_{\mathrm{inv}}$ in pp collisions at $\sqrt{s}=0.9$ TeV as function of pair transverse momentum \cite{Aamodt:2010jj}.}
\label{fig:twopion2}
\end{minipage}
\end{figure}

In agreement with other measurements of hadron-hadron collisions above $\sqrt{s}\sim 50$~GeV, ALICE has measured an increase of the HBT radius $R_{\mathrm{inv}}$ with multiplicity (Figure \ref{fig:twopion}). 
On the other hand, the $\langle k_{\mathrm{T}} \rangle$ dependence of the measured HBT radius of ALICE (Figure \ref{fig:twopion2}) 
exhibits a different behavior as compared to other experiments at lower $\sqrt{s}$:
ALICE has measured an $R_{\mathrm{inv}}$, which is practically independent of $\langle k_{\mathrm{T}} \rangle$ within the studied $\langle k_{\mathrm{T}} \rangle$ range whereas STAR and E735 have observed a decrease of $R_{\mathrm{inv}}$ with increasing $\langle k_{\mathrm{T}} \rangle$.
Here, it has to be taken into account that this measurement is sensitive to the choice of the baseline. A usage of a flat baseline changes the $\langle k_{\mathrm{T}} \rangle$ dependence to a falling $R_{\mathrm{inv}}$ with increasing $\langle k_{\mathrm{T}} \rangle$. STAR and E735 have used a flat baseline. 

\section{Mid-rapidity antiproton-to-proton ratio}
The initial state of a pp collision has a total baryon number of 2. Studies of the redistribution of the baryon number in the final state allow for investigation of the baryon number transfer and its reach to mid-rapidity. The rapidity loss of a baryon originating in the beam is given by $\Delta y = y_{\mathrm{beam}}-y_{\mathrm{baryon}}$.\\
The baryon number transfer can be modeled by the breaking of strings between the valence quarks and so-called string junctions. Processes with large $\Delta y$ can be described by Regge trajectories.\\
In order to investigate the baryon number transfer, ALICE has measured the \={p}/p ratio at mid-rapidity at $\sqrt{s}$ = 0.9 and 7 TeV \cite{Aamodt:2010dx}. Most of the p and \={p} at mid-rapidity are created in baryon pair-production resulting in equal yields. Any excess of p over \={p} is therefore associated to a baryon number transfer. For this challenging measurement, very precise knowledge of the material budget and the cross sections is required.\\% The measurement was performed as function of transverse momentum and as function of the collision energy. \\
\vspace{-0.8cm}
\begin{figure}[ht]
\centering
\includegraphics[width=0.48\textwidth]{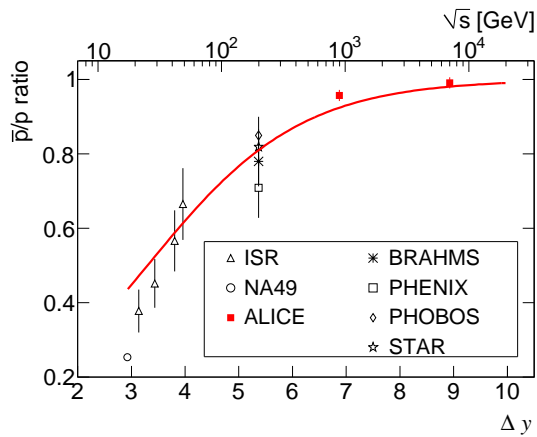}
\vspace{-0.5cm}
\caption{Central rapidity \={p}/p ratio as a function
of the rapidity interval $\Delta y$ (lower axis) and center-of-mass
energy (upper axis) \cite{Aamodt:2010dx}.}
\label{fig:ppbar}
\end{figure}

Results at both energies show no dependence on the transverse momentum (plot not shown)\cite{Aamodt:2010dx}. The energy dependence of the transverse momentum integrated ratio presented in Figure \ref{fig:ppbar} can be parameterized based on the contribution of different diagrams including baryon pair production at mid-rapidity and baryon number transfer. A fit with a junction intercept set to 0.5 reproduces the distribution well. The presented result sets tight limits on any additional contributions to baryon number transfer over large rapidity gaps.

\section{Production of pions, kaons and protons}
\begin{figure}[ht]
\begin{minipage}[b]{0.4\linewidth}
\centering
\includegraphics[width=\textwidth]{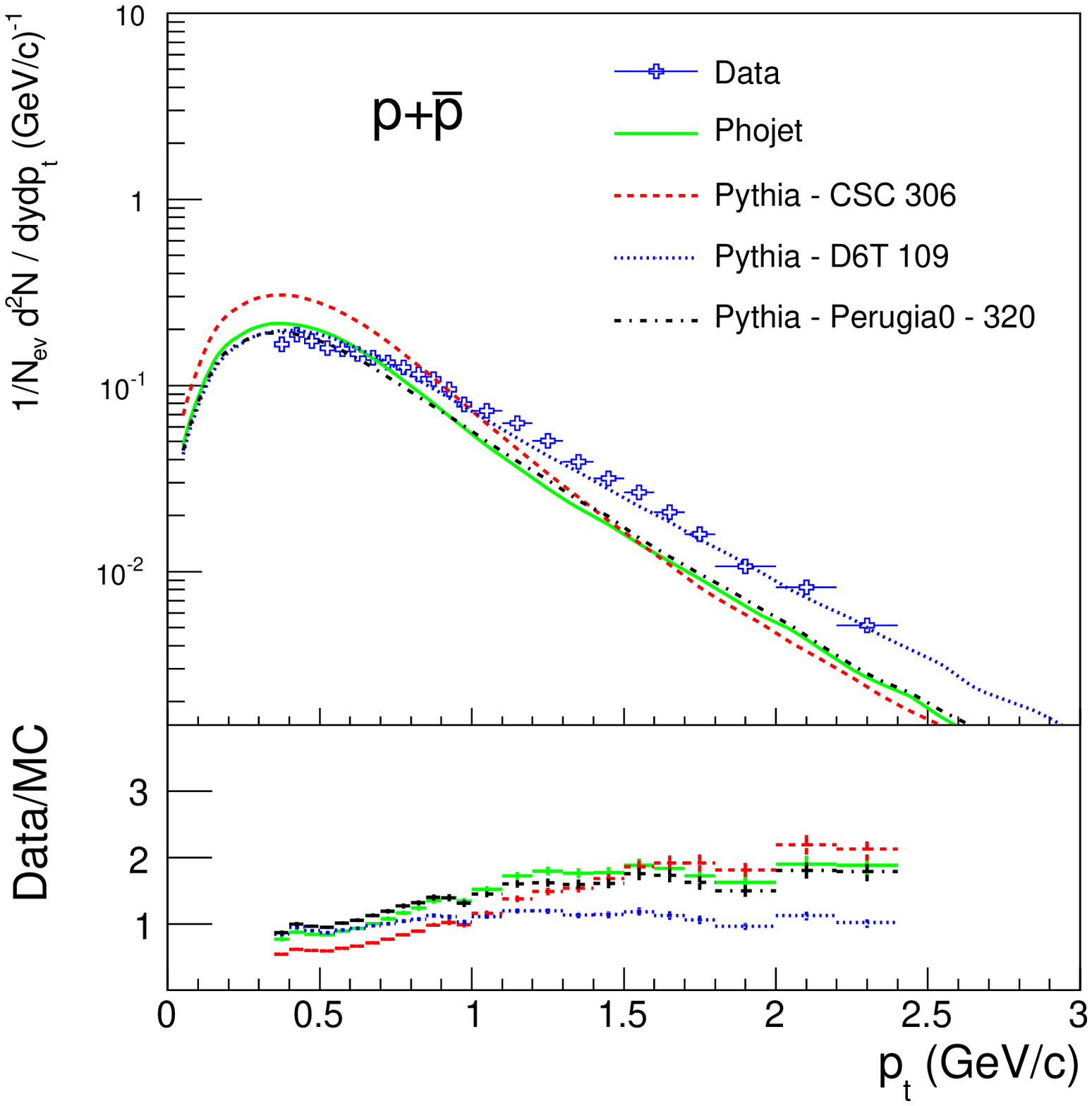}
\end{minipage}
\vspace{-0.5cm}
\hspace{2.8cm}
\begin{minipage}[b]{0.39\linewidth}
\flushright
\includegraphics[width=\textwidth]{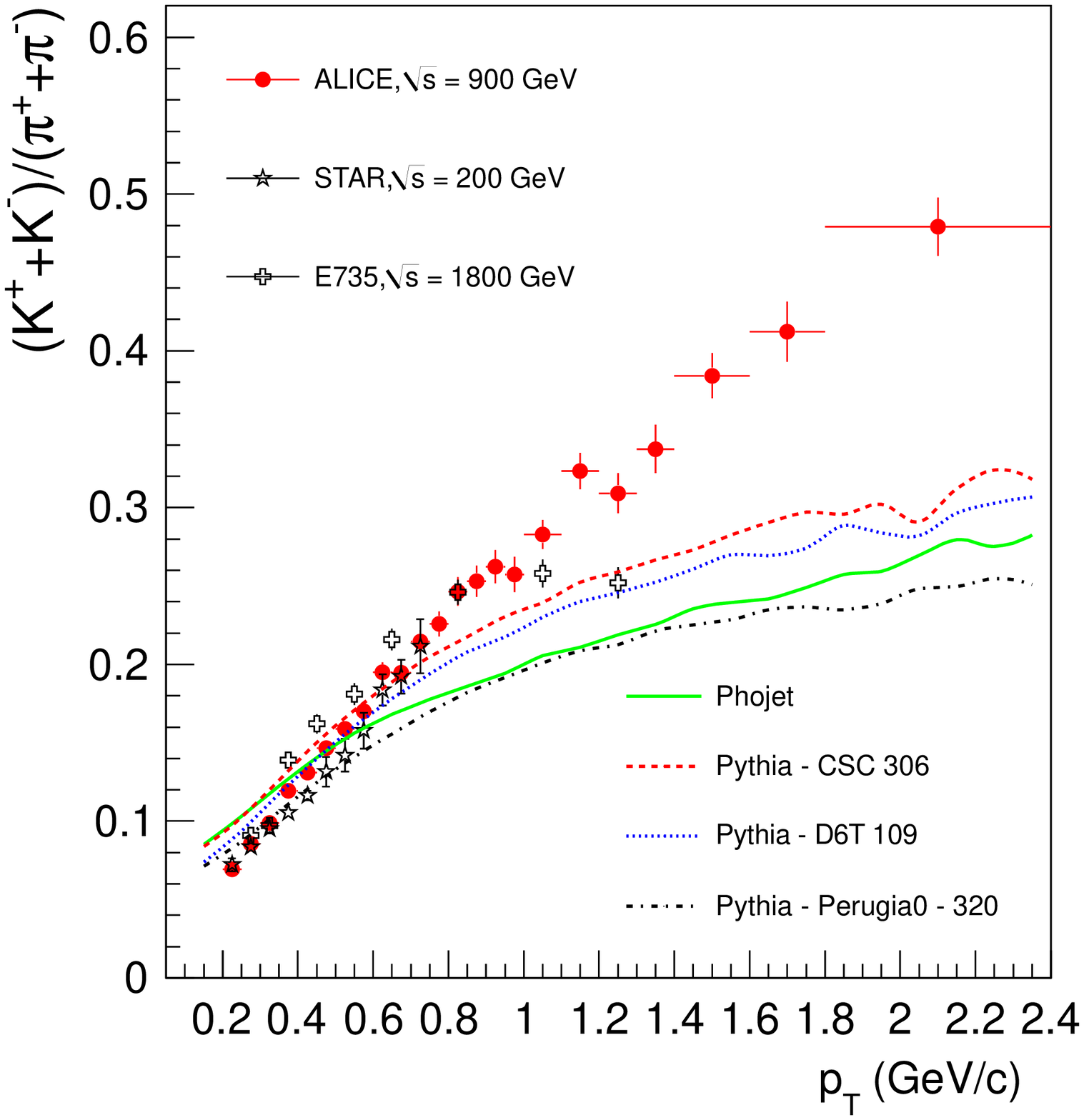}
\end{minipage}
\vspace{-0.5cm}
\begin{minipage}[t]{0.45\linewidth}
\centering
\caption{Comparison of the p+\={p} spectrum measured at $\sqrt{s}$ = 900 GeV and model predictions \cite{Aamodt:2011zj}.}
\label{fig:id}
\end{minipage}
\hspace{1.2cm}
\begin{minipage}[t]{0.45\linewidth}
\centering
\caption{Ratio of kaon to pion spectra as a function of the transverse momentum \cite{Aamodt:2011zj}.}
\label{fig:id2}
\end{minipage}
\end{figure}

ALICE identifies $\pi^{\pm}$, $K^{\pm}$, p, and \={p} using different methods such as specific energy loss in the Inner Tracking System (ITS) and the Time Projection Chamber (TPC) and the particle's time of flight in the dedicated Time Of Flight (TOF) detector \cite{Aamodt:2011zj}. The identification can be performed on a track-by-track basis where the bands are clearly separated as well as on a statistical basis in the overlapping areas. For kaons, a complementary measurement can be performed via the identification of their weak decay kink topology in the TPC.\\
The detectors and measurement approaches cover different momentum ranges. Where the transverse momentum ranges overlap, the results are in good agreement. The combined spectra are estimated by averaging the sub-detector's results using the systematic errors as weight.\\
A fit using the so-called L\'{e}vy function yields a good description of the spectra. As an example, Figure \ref{fig:id} presents the combined transverse momentum spectrum for protons and anti-protons. The PYTHIA tunes Perugia-0 and D6T gave a reasonable description of the unidentified charged hadron spectra \cite{Aamodt:2010jj}. The spectra of identified hadrons, however, especially the kaon and proton spectra, show large deviations: The kaon yield is underestimated at high $p_{\mathrm{T}}$ by all models (plot not shown). The proton yield can only be described by PYTHIA D6T. \\
The kaon to pion ratio visible in Figure \ref{fig:id2} is almost independent of the center-of-mass energy. The ratio is underestimated by all models at high $p_{\mathrm{T}}$.

\section{Strange particle production}
\begin{figure}[ht]
\begin{minipage}[b]{0.4\linewidth}
\centering
\includegraphics[width=\textwidth]{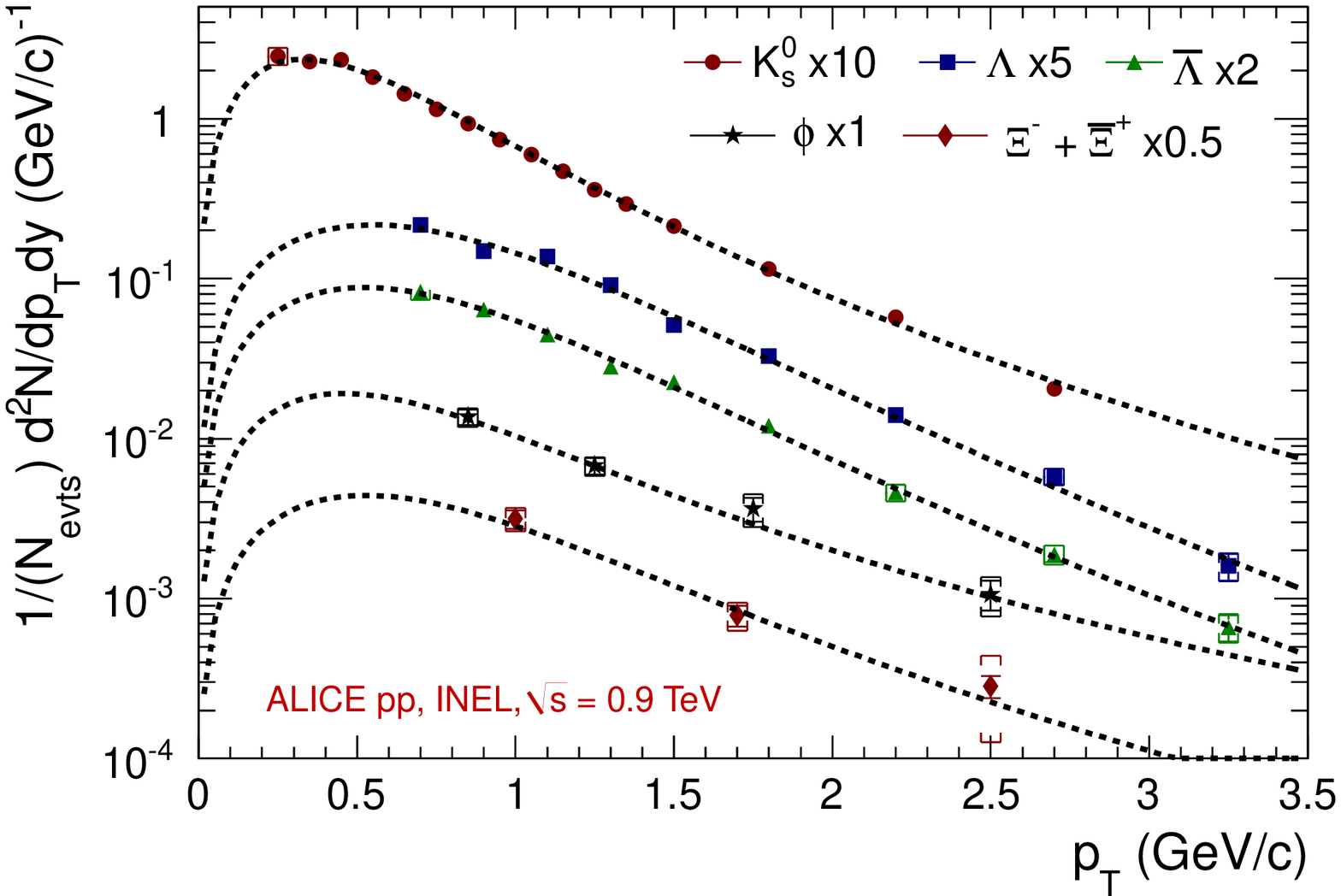}
\end{minipage}
\hspace{2.9cm}
\begin{minipage}[b]{0.4\linewidth}
\flushright
\includegraphics[width=\textwidth]{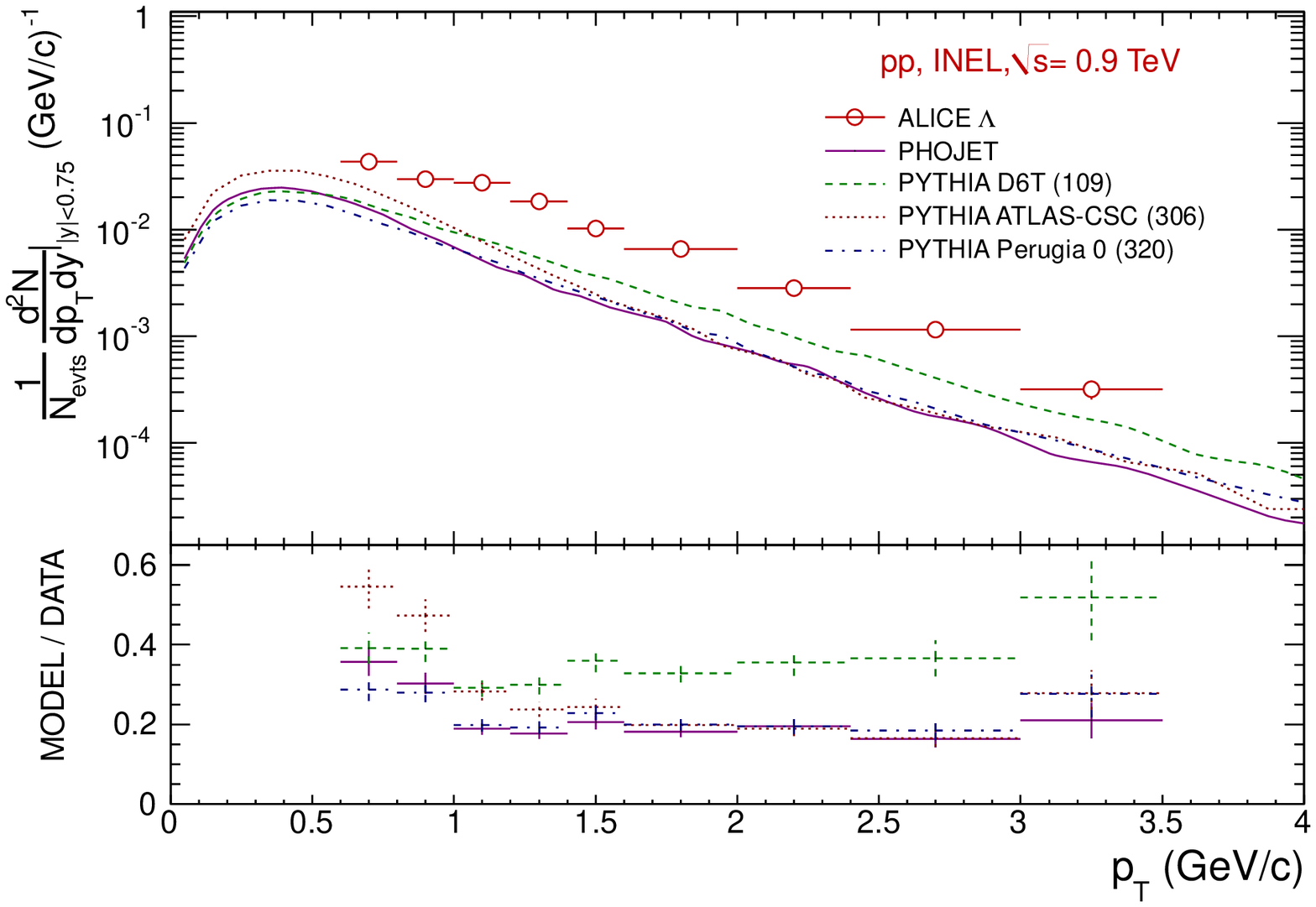}
\end{minipage}

\begin{minipage}[t]{0.45\linewidth}
\centering
\caption{Particle spectra as a function of $p_{\mathrm{T}}$ for  $K^0_s$, $\Phi$, $\Lambda$ and $\Xi^-$ and $\bar{\Xi}^+$ \cite{Aamodt:2011zz}. The data points are scaled for better visibility.}
\label{fig:strange}
\end{minipage}
\hspace{1.2cm}
\begin{minipage}[t]{0.45\linewidth}
\centering
\caption{Comparison of the transverse momentum differential yield for the $\Lambda$ particles for INEL pp collisions with models \cite{Aamodt:2011zz}.}
\label{fig:strange2}
\end{minipage}
\end{figure}

ALICE has measured the production of strange mesons and single and double strange baryons at central rapidity at $\sqrt{s}$ = 900 GeV \cite{Aamodt:2011zz}. Figure \ref{fig:strange} shows the transverse momentum spectra for $K_\mathrm{s}^0$, $\Phi$, $\Lambda$ and $\Xi^{-}+\bar{\Xi}^+$ scaled for visibility and fitted by the above mentioned L\'{e}vy function. \\
As an example for all spectra, the spectrum of $\Lambda$ particles is presented in Figure \ref{fig:strange2}. For all species, the $p_{\mathrm{T}}$ spectra are found to be slightly harder than predicted by any of the tested event generators. For $p_{\mathrm{T}}>1$ GeV/$c$, the strange particle spectra are strongly underestimated, with the exception of the $\Phi$(= s \=s) which is well described.

\section{Charged-particle multiplicity density in central Pb--Pb collisions}
One of the first measurements of ALICE was the measurement of the charged-particle multiplicity density at mid-rapidity in Pb--Pb collisions at a center of mass energy of $\sqrt{s_{\mathrm{NN}}}$ = 2.76 TeV for the most central 5\% of the hadronic cross section. 
This measurement is shown in Figure 13 in conjunction with measurements in pp and A--A at smaller $\sqrt{s}$.
We observe a stronger energy dependence in Pb--Pb as compared to pp and an increase by a factor of 1.9 with respect to pp collisions at similar energies. The values are significantly larger than those measured at RHIC, namely there is an increase by a factor of 2.2 with respect to RHIC Au--Au at 200 GeV.

\begin{figure}[ht]
\begin{minipage}[t]{0.4\linewidth}
\centering
\includegraphics[width=\textwidth]{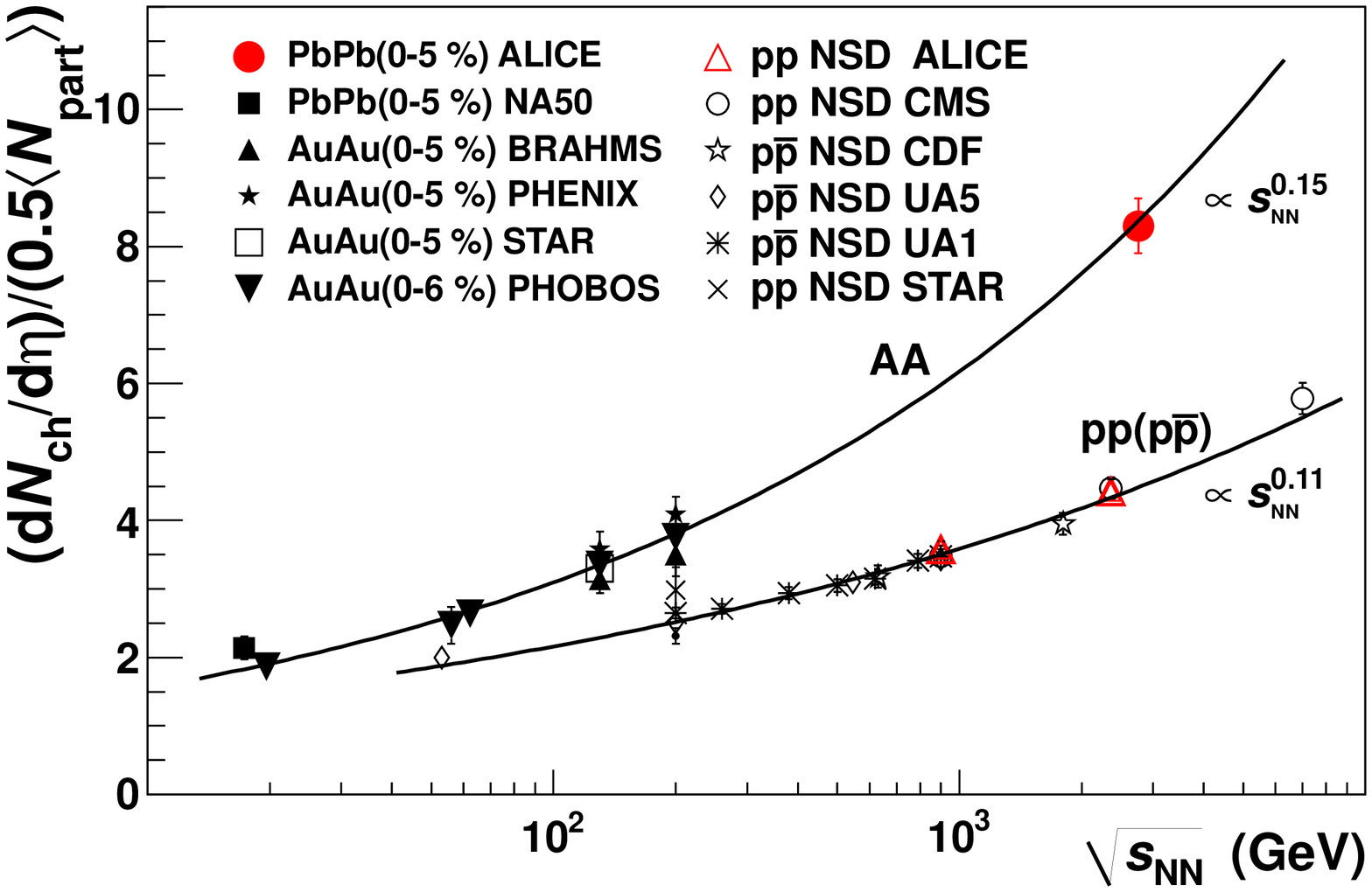}
\end{minipage}
\hspace{3.0cm}
\begin{minipage}[t]{0.35\linewidth}
\flushright
\includegraphics[width=\textwidth]{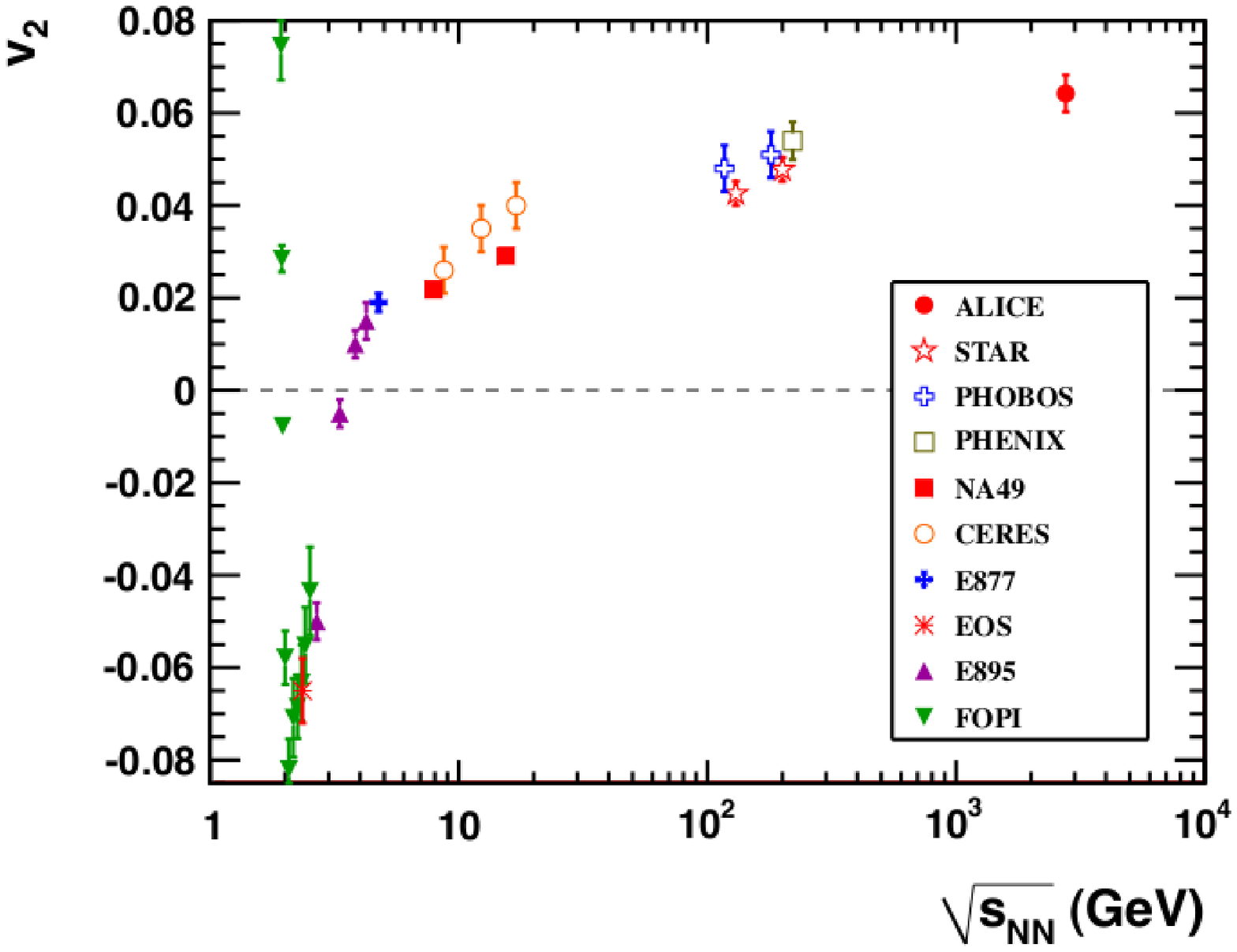}
\end{minipage}

\begin{minipage}[t]{0.45\linewidth}
\centering
\caption{Charged particle pseudorapidity density per participant pair for central nucleus-nucleus and non-single diffractive pp (p\={p}) collisions as a function of $\sqrt{s_{\mathrm{NN}}}$ \cite{Aamodt:2010pb}.}
\label{fig:pbpb}
\end{minipage}
\hspace{1.2cm}
\begin{minipage}[t]{0.45\linewidth}
\centering
\caption{Integrated elliptic flow at 2.76 TeV in Pb--Pb 20--30\% centrality compared to results at similar centralities \cite{Aamodt:2010pa}.}
\label{fig:pbpb2}
\end{minipage}
\end{figure}

\section{Elliptic flow of charged particles in Pb--Pb collisions}
The overlap zone in a non-central nucleus-nucleus collision is asymmetric. If the matter in the collision region is interacting, the spacial asymmetry is converted into an anisotropic momentum distribution owing to different pressure gradients. The second Fourier coefficient of the resulting final state hadron azimuthal distribution is call elliptic flow ($v_2$). \\
ALICE has measured the integrated elliptic flow $v_2$ at 20--30\% centrality \cite{Aamodt:2010pa}, which is about 30\% larger at the LHC as compared to RHIC (Figure \ref{fig:pbpb2}). The increase is larger than predictions from ideal hydrodynamic models. However, hydrodynamical models that incorporate viscous corrections and hybrid models are able to reproduce the observed increase.

\section{Summary}
After about one year of LHC data taking, ALICE has measured a comprehensive list of minimum bias observables in pp and in Pb--Pb collision events. Further analyses are progressing well, especially those that rely on a larger sample of collision data or rare triggers.

\end{document}

%% file: econfmacros.tex
\def\beq{\begin{equation}}
\def\eeq#1{\label{#1}\end{equation}}
\def\eeqn{\end{equation}}

\def\beqa{\begin{eqnarray}}
\def\eeqa#1{\label{#1}\end{eqnarray}}
\def\eeqan{\end{eqnarray}}

\let\bar=\overbar

\def\Dslash{\not{\hbox{\kern-4pt $D$}}}
\def\dslash{\not{\hbox{\kern-2pt $\del$}}}

\def\msb{{\bar{\ssstyle M \kern -1pt S}}}